\documentstyle[twoside,fleqn,espcrc2,epsf]{article}
\newcommand{\ovl}[1]{\overline{#1}}

\newcommand{\re}{{\rm Re}}
\newcommand{\im}{{\rm Im}}

\newcommand{\VEV}[3]{\left\langle #1\left| #2 \right| #3\right\rangle}

\title{ Calculation of $K\to\pi\pi$ decay amplitudes from $K\to\pi$
        matrix elements in quenched domain-wall QCD
\thanks{Talk presented by J.~Noaki} }

\author{CP-PACS Collaboration :
  S.~Aoki\rlap,\address{Institute of Physics, University of Tsukuba, 
  Tsukuba, Ibaraki 305-8571, Japan}
  Y.~Aoki\rlap,\address{Center for Computational Physics,
    University of Tsukuba, Tsukuba, Ibaraki 305-8577,
    Japan}\thanks{present address: RIKEN BNL Research Center, 
    Brookhaven National Laboratory, Upton, NY 11973, USA}
  R.~Burkhalter\rlap,$^{\rm b}$
  S.~Ejiri\rlap,$^{\rm b}$\thanks{present address: Department of
  Physics, University of Wales Swansea, Singleton Park, Swansea SA2 8PP, UK},
  M.~Fukugita\rlap,\address{Institute for Cosmic Ray Research,
    University of Tokyo, Kashiwa 277-8582, Japan}
  S.~Hashimoto\rlap,\address{High Energy Accelerator Research Organization
    (KEK), Tsukuba, Ibaraki 305-0801, Japan}
  N.~Ishizuka\rlap,$^{\rm a,b}$
  Y.~Iwasaki\rlap,$^{\rm a,b}$
  T.~Izubuchi\rlap,\address{Institute of Theoretical Physics, Kanazawa
    University, Ishikawa 920-1192, Japan}\thanks{presently on leave at :
  Physics Depertment, Brookhaven National Laboratory, Upton, NY 11973, USA}
  K.~Kanaya\rlap,$^{\rm a}$
  T.~Kaneko\rlap,$^{\rm e}$
  Y.~Kuramashi\rlap,$^{\rm e}$
  V.~Lesk\rlap,$^{\rm b}$
  K.-I.~Nagai\rlap,$^{\rm b}$\thanks{present address: Theory Division, CERN, 
CH-1211 Geneva 23, Switzerland}
  J.~Noaki\rlap,$^{\rm a \dagger}$
  M.~Okawa\rlap,$^{\rm e}$
  Y.~Taniguchi\rlap,$^{\rm a}$
  A.~Ukawa$^{\rm a,b}$ and
  T.~Yoshi\'e$^{\rm a,b}$
  }

\begin{document}

\begin{abstract}
We present a calculation of the $K\to\pi\pi$ decay amplitudes 
from the $K\to\pi$ matrix elements using leading order relations 
derived in chiral perturbation theory. 
Numerical simulations are carried out in quenched QCD with the 
domain-wall fermion action and the renormalization group improved 
gluon action.  Our results show that the $I=2$ amplitude is 
reasonably consistent with experiment whereas the $I=0$ amplitude 
is sizably smaller.  Consequently the $\Delta I=1/2$ enhancement is 
only half of the experimental value, 
and $\varepsilon'/\varepsilon$ is negative.
\end{abstract}
\maketitle

\section{INTRODUCTION}
Quantitative understanding of the $K\to\pi\pi$ decay including 
the $\Delta I=1/2$ rule and the value of $\varepsilon'/\varepsilon$ 
has been a long-standing issue in lattice QCD~\cite{Martinelli}. 
Here we present a summary of our work~\cite{KPPpaper} based on 
(i) the leading order chiral perturbation theory (ChPT) formula 
relating the $K\to\pi\pi$ decay amplitudes to the $K\to\pi$ matrix 
elements~\cite{Bernard85}, (ii) domain-wall fermion action 
to ensure chiral symmetry at finite lattice spacings in principle, 
and (iii) the renormalization group (RG) improved gauge action 
for suppression of chiral breaking effects for moderate fifth 
dimensional lattice size over those for the standard plaquette 
gauge action~\cite{CPPACS_origA}.  
For a similar work using the plaquette gauge action 
we refer to Ref.~\cite{Columbia}.  

\section{NUMERICAL SIMULATION}

\begin{table}
  \caption{Numbers of configurations generated in our calculation
 for each combination of $m_f$ and volume.}
 \begin{tabular}{|c|ccccc|}
  \hline
  $m_f$ & 0.02 & 0.03 & 0.04 & 0.05 & 0.06 \\
  \hline
  $16^3\times 32$ & 407 & 406 & 406 & 432 & 435 \\
  $24^3\times 32$ & 432 & 200 & 200 & 200 & 200 \\
  \hline
  \end{tabular}
 \label{confs-table}
\vspace{-0.7cm}
\end{table}

The numerical task at hand is to calculate the matrix elements 
$\VEV{\pi^+}{Q_i^{(I)}}{K^+}$ and $\VEV{0}{Q_i^{(0)}}{K^0}$
where $Q_i^{(I)} (i=1, \cdots, 10) $ are the local 4-quark operators 
of isospin $I=0, 2$ that appear in the effective weak Hamiltonian. 

This calculation is carried out with the RG-improved gauge action 
at $\beta=2.6$, which corresponds to the scale $1/a=1.94$ GeV 
determined from the string tension $\sqrt{\sigma}=440$ MeV. 
For the domain-wall fermion action, we use the domain wall height 
$M=1.8$ and the fifth dimensional size $N_s=16$. We consider the case of 
degenerate $u$-$d$ and $s$ quark with a common mass $m_f$.   
Gauge configurations are generated {\it independently} 
for each value of $m_f$.  
Our statistics are summarized in Table~\ref{confs-table} for the two 
lattice volumes and $m_f$ used in the calculation. 

\section{CHIRAL PROPERTIES}
\begin{figure}[t]
 \vspace*{-0.43cm}
 \begin{center}
  \leavevmode
  \hspace*{-0.2cm}
  \epsfxsize=7.9cm \epsfbox{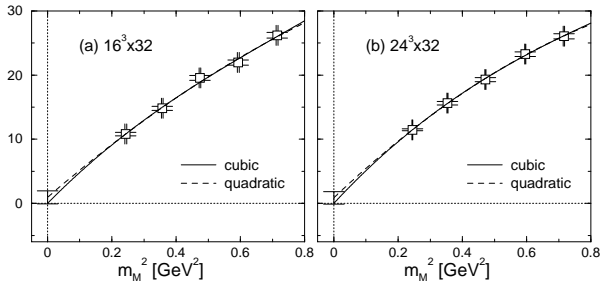}
 \end{center}
 \vspace{-1.4cm}
  \caption{$K^+\to\pi^+$ matrix element of $Q_6^{(0)}$ 
 for (a) $16^3$ and (b) $24^3$. Solid and dashed curves show 
 chiral fits as described in text. }
  \label{Q06_chiprop}
 \vspace{-0.6cm}
\end{figure}

\begin{figure}[h]
 \vspace*{-0.4cm}
  \begin{center}
   \leavevmode
  \hspace*{-0.6cm}
  \epsfxsize=7.9cm \epsfbox{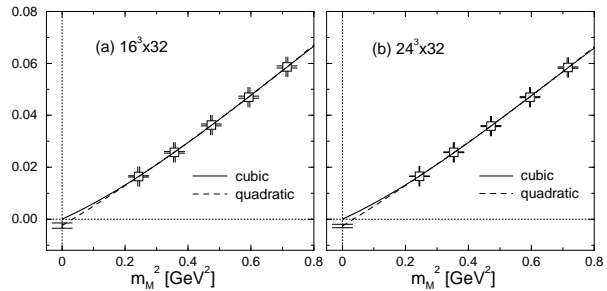}
  \end{center}
 \vspace{-1.4cm}
  \caption{Same as Fig.~\ref{Q06_chiprop} but for $Q_1^{(2)}$.}
  \label{Q21_chiprop}
 \vspace{-0.5cm}
\end{figure}

Our representative results for the $K\to\pi$ matrix elements are 
shown in Figs.~\ref{Q06_chiprop} and~\ref{Q21_chiprop}.  
In terms of the lattice pion mass squared, $m_M^2$, the $K\to\pi$ 
matrix elements are expanded as 
\begin{eqnarray}
\VEV{\pi^+}{Q_i^{(I)}}{K^+}=a_0+a_1m_M^2+a_2(m_M^2)^2\nonumber \\ 
+a_3(m_M^2)^2\ln m_M^2+ a_4(m_M^2)^3+\cdots.
\end{eqnarray}
Checking the chiral property with a quadratic fit 
with the fit parameters $(a_0,a_1,a_2)$, we find the constant $a_0$ 
to be consistent with zero within statistical errors for most 
of matrix elements.  The only exception is $Q_1^{(2)}$ as seen in 
Fig.~\ref{Q21_chiprop}.  However, fits with the constraint $a_0=0$ 
such as a cubic one using $(a_1,a_2,a_4)$ or a form with the logarithm 
$(a_1,a_2,a_3)$ yield acceptable $\chi^2$/DOF. 
We therefore consider that the chiral properties required for viability 
of the ChPT relations between the $K\to\pi\pi$ and $K\to\pi$ 
amplitudes hold well.  

An open problem is a small value of the coefficient of the 
chiral logarithm term obtained in the fits, {\it e.g.,} 
$a_3/a_1= -0.58\ {\rm GeV^{-2}}$ for $Q_1^{(2)}$, which is only 
about 25\% of the prediction of chiral perturbation 
theory~\cite{Gol-Palla}: $-3/(8\pi^2f_\pi^2)$.

\section{PHYSICAL RESULTS}

We construct the $K\to\pi\pi$ decay amplitudes from 
the $K\to\pi$ matrix elements as described in Ref.~\cite{KPPpaper}.  
Renormalization of the lattice matrix 
elements is made with tadpole-improved perturbation theory at one-loop 
level~\cite{renorm} at the matching point $q^*=1/a$. 
The renormalization group running from $\mu=1/a$ to $m_c$ (=1.3 GeV) 
is carried out for the $N_f=3$ theory with $\Lambda_{\ovl{\rm MS}}=325$ 
MeV~\cite{Buch-Buras-Lauten95}. 
Combining these results with the Wilson coefficients lead to the physical  
results for the $K\to\pi\pi$ decay amplitudes $A_I\ (I=0,2)$.

\begin{figure*}[t]
 \vspace*{-0.4cm}
 \begin{center}
  \leavevmode
  \epsfxsize=4.9cm \epsfbox{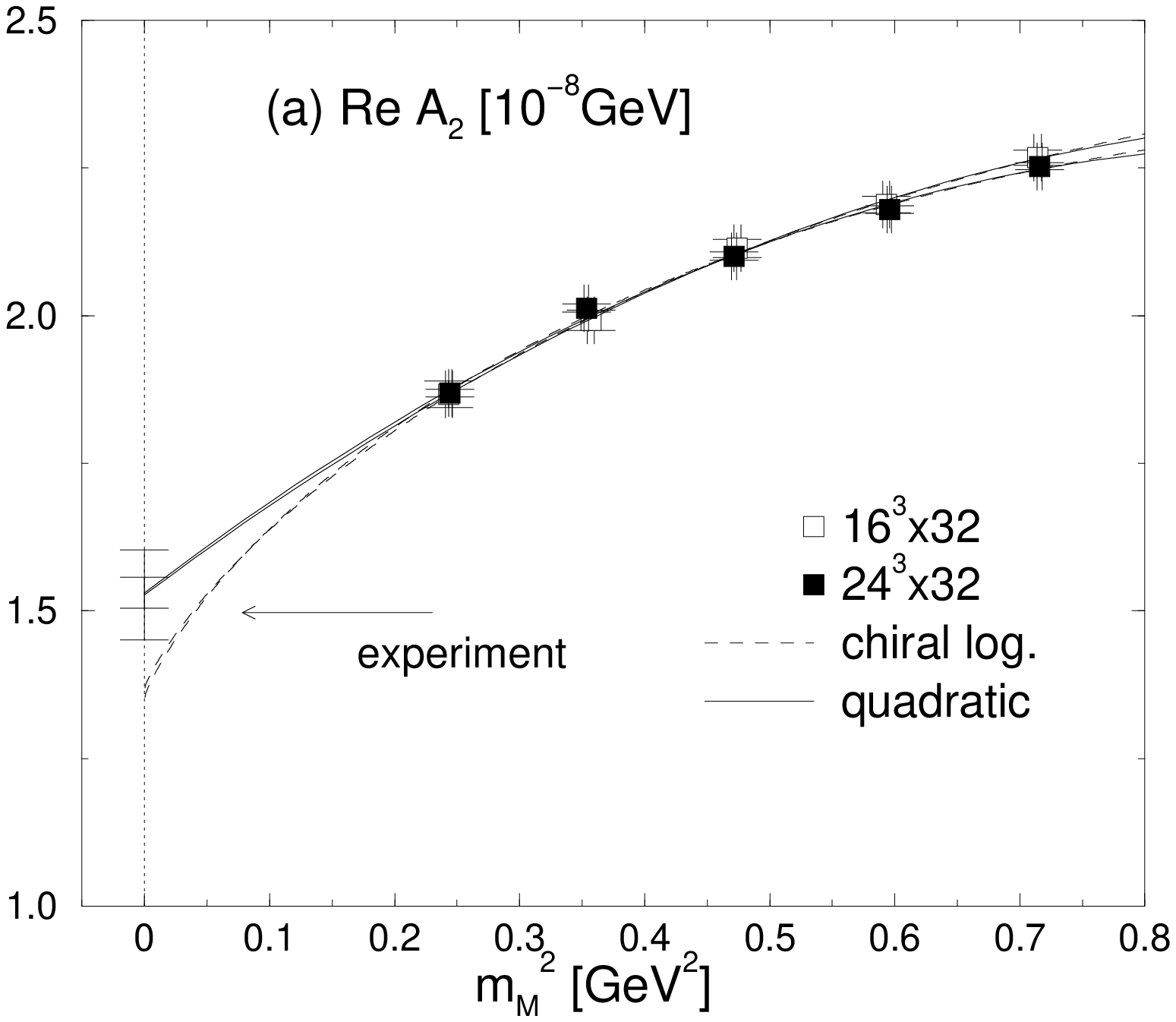}
  \epsfxsize=4.9cm \epsfbox{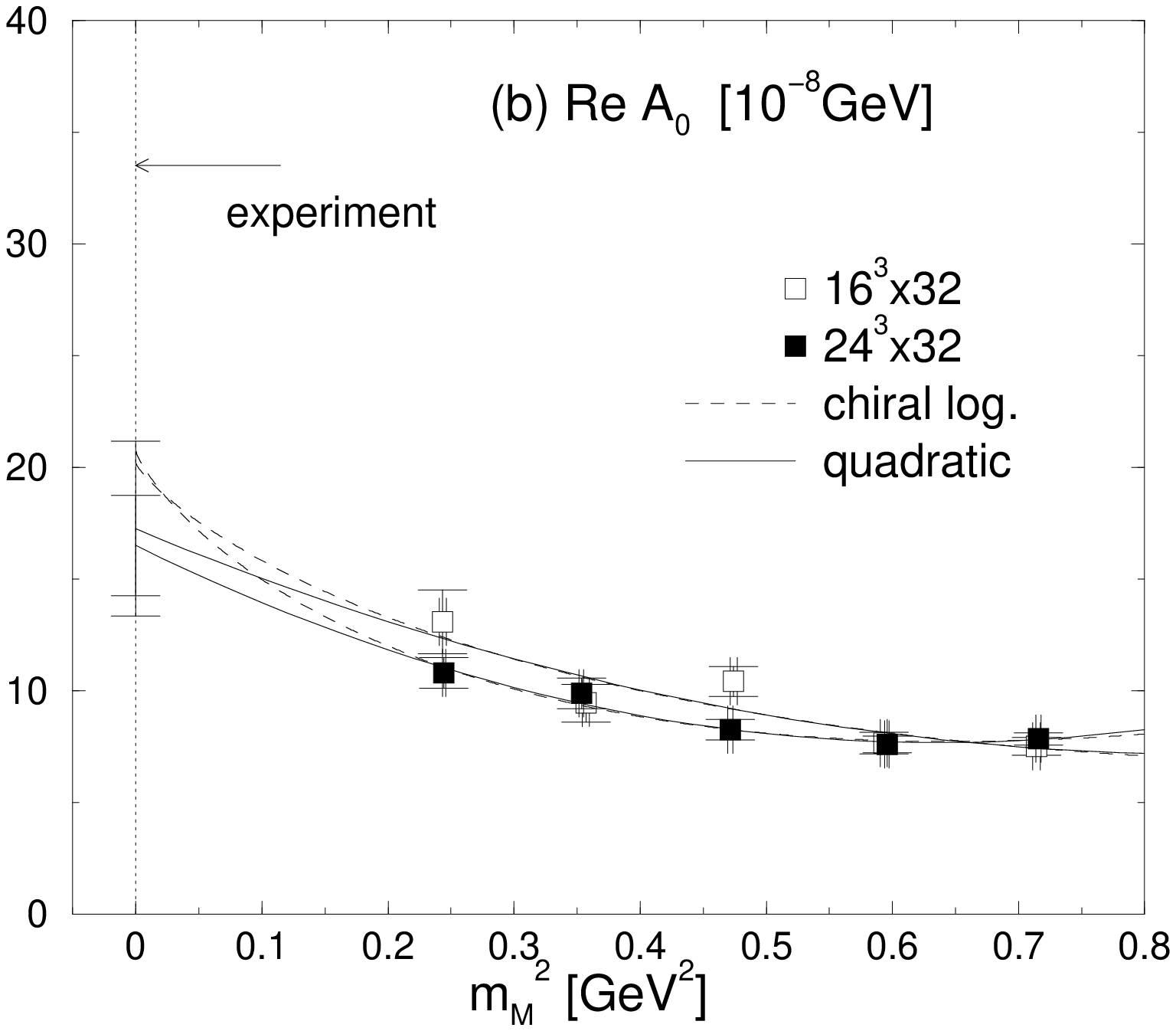}
  \epsfxsize=4.9cm \epsfbox{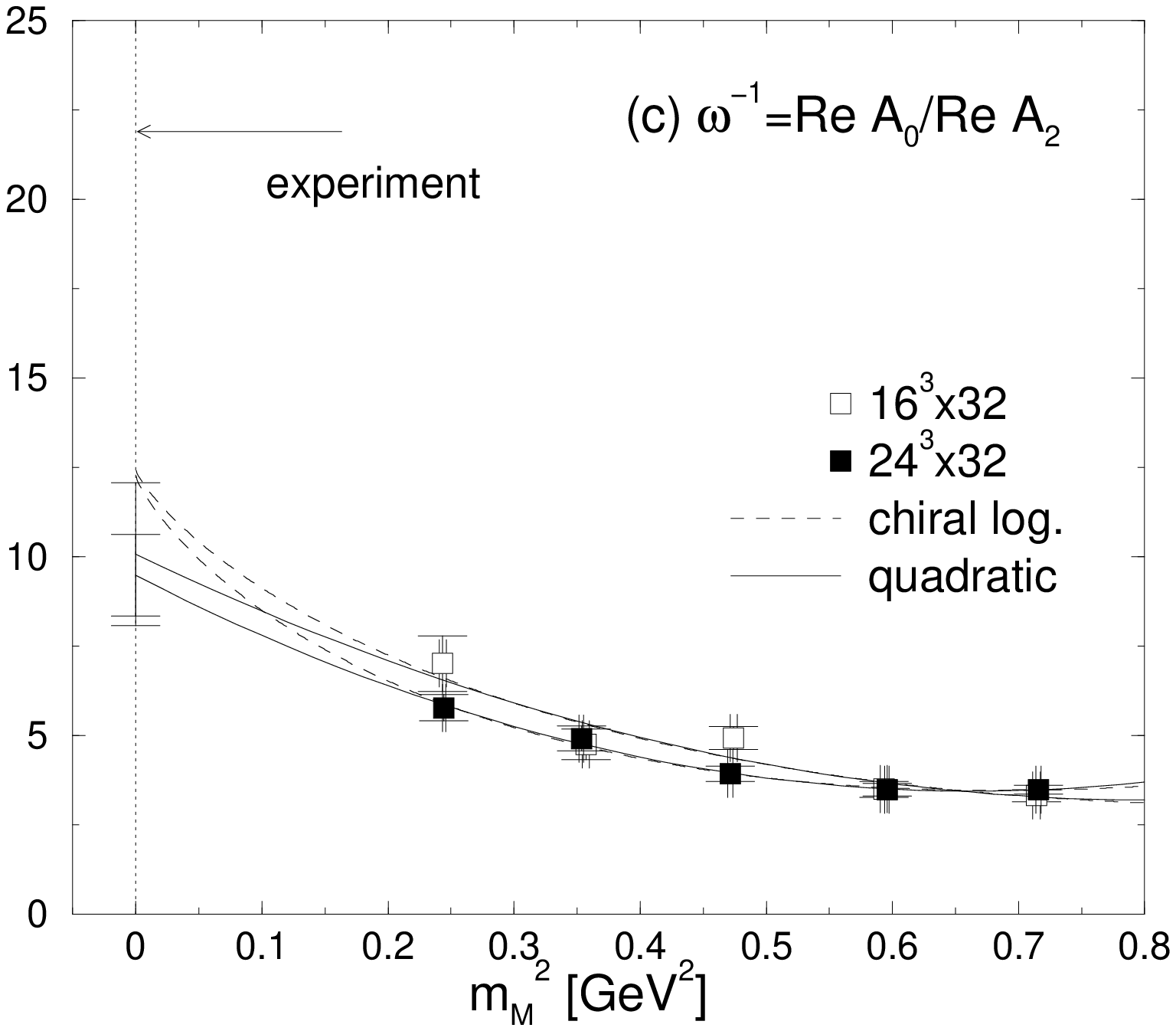}
 \end{center}
 \vspace{-1.4cm}
 \caption{Real parts of the decay amplitude $\re A_I\ [10^{-8}{\rm GeV}] $ 
 and the ratio $\omega^{-1}=\re A_0/\re A_2$ as a function of 
 $m_M^2\ [{\rm GeV}^2]$. Open (filled) symbols are from the volume 
 $16^3\ (24^3)$, and solid (dashed) curve show quadratic (chiral logarithm) 
 fit result.}
 \label{deltaI-fig}
\end{figure*}

In Fig.~\ref{deltaI-fig} we present $\re A_2$, $\re A_0$ and 
$\omega^{-1}=\re A_0/\re A_2$ as a function of $m_M^2$. 
We observe that the data from the volumes $16^3$ 
(open symbols) and $24^3$ (filled symbols) are consistent.
Lines drawn are the fit curves for the quadratic (solid) function in 
$m_M^2$ or a form including chiral logarithm (dashed). 
The $I=0$ amplitude reaches only half of the experimental value in the 
chiral limit, and hence the $\Delta I=1/2$ rule, {\it i.e.,} 
$\omega^{-1}\approx 22$, is reproduced only by about 50\% in the chiral limit. 

\begin{figure*}[t]
 \vspace*{-0.4cm}
  \begin{center}
   \leavevmode
   \epsfxsize=4.7cm \epsfbox{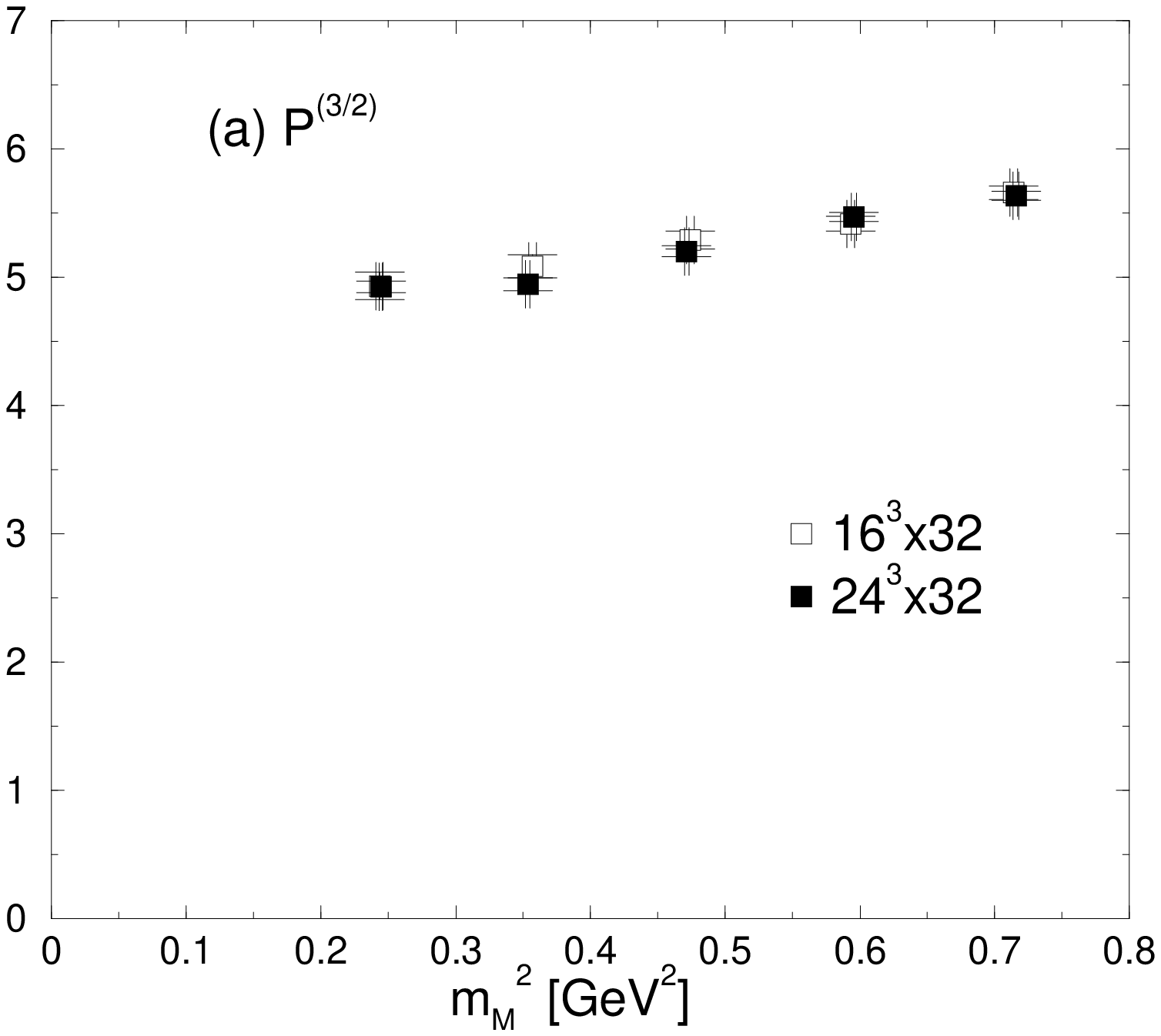}
   \hspace{0.05cm}
   \epsfxsize=4.7cm \epsfbox{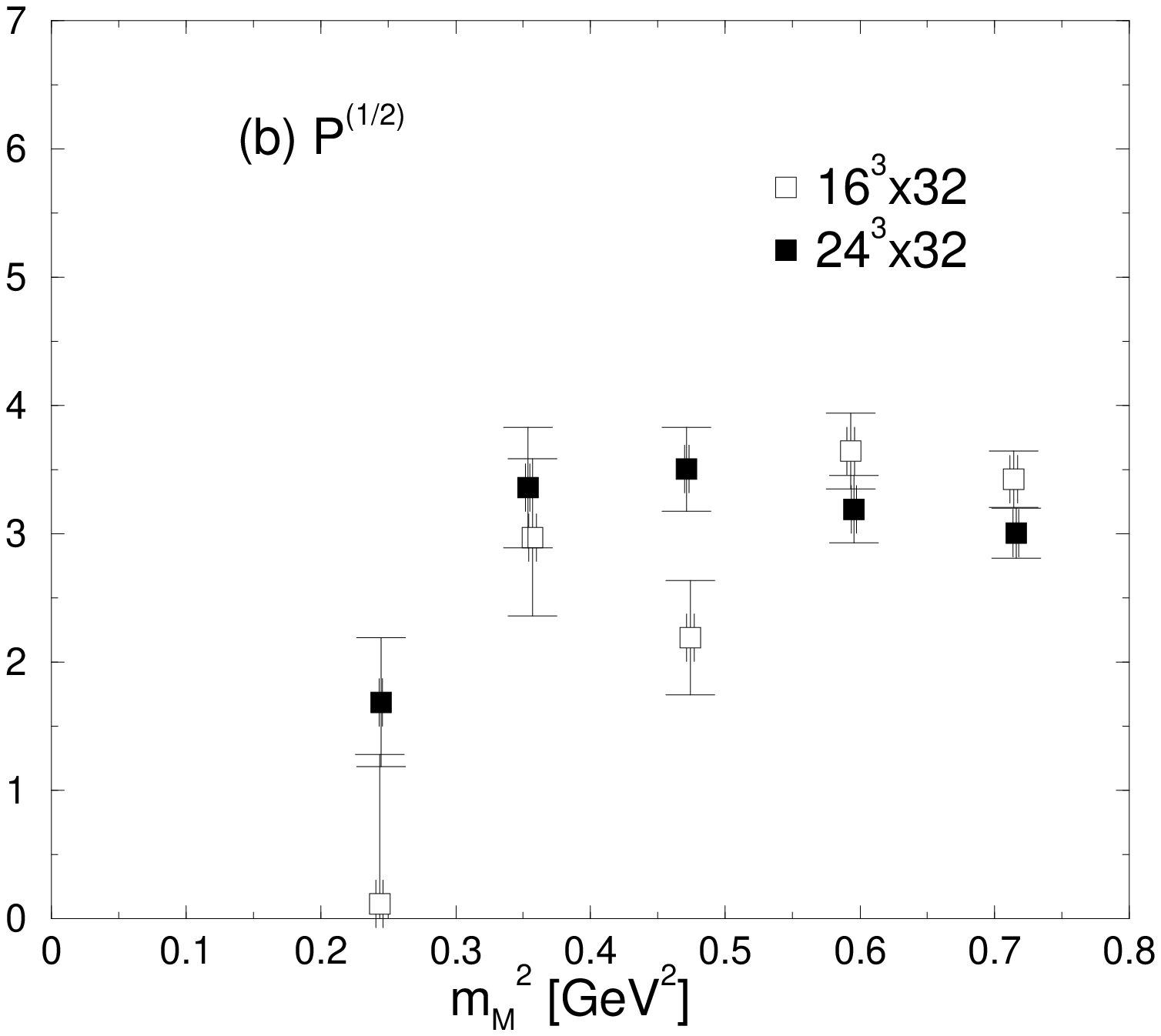}
   \epsfxsize=4.9cm \epsfbox{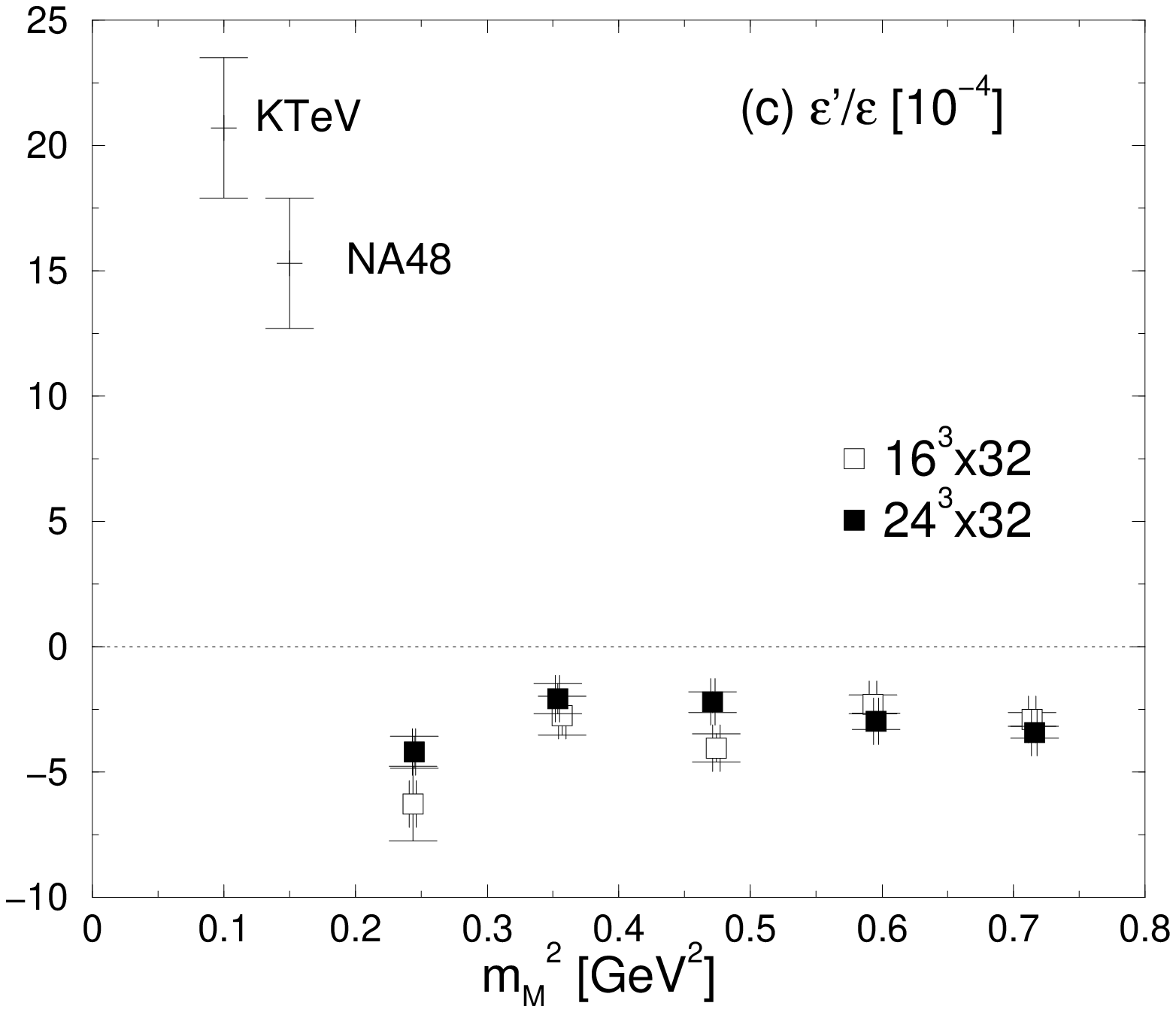}
  \end{center}
 \vspace{-1.4cm}
 \caption{Imaginary parts of the decay amplitude $P^{(3/2)}$, $P^{(1/2)}$ and 
  $\varepsilon'/\varepsilon$.
  Meaning of symbols is same as in Fig.~\ref{deltaI-fig}.}
 \label{epep-fig}
 \vspace*{-0.35cm}
\end{figure*}

In Fig.~\ref{epep-fig} we plot $P^{(3/2)}$ and 
$P^{(1/2)}$, which are related to $\varepsilon'/\varepsilon$ 
as
~\cite{Buch-Buras-Lauten95} as
\begin{eqnarray}
\varepsilon'/\varepsilon = \im (V_{\rm ts}^*V_{\rm td})[ P^{(1/2)}-P^{(3/2)}].
\end{eqnarray}
Here we use our results only for the factor $\im A_{0,2}$. 
substituting the experimental value for the real part of the amplitudes. 
Our results for $\varepsilon'/\varepsilon$ is negative since 
$P^{(3/2)}>P^{(1/2)}$. 

\section{PENGUIN OPERATORS}

Recently, Golterman and Pallante~\cite{GOL-PAL} pointed out
a necessity of new operators to represent the penguin operators 
in (partially) quenched chiral perturbation theory,
and gave a simple prescription to extract the physical 
contribution of $Q_{5,6}$ to the $K\to\pi$ and $K\to 0$ matrix elements
in this framework.
Our preliminary analyses show that this modification, if employed,
increases $P^{(1/2)}$ by about 10--50\% depending on $m_f$, but 
this effect is too small to affect the negative value of 
$\varepsilon^\prime/\epsilon$ we have obtained. 

\vspace*{3mm}
This work is supported in part by Grants-in-
Aid of the Ministry of Education (Nos. 10640246,\\
10640248, 10740107, 11640250, 11640294,\\ 
11740162, 12014202, 12304011, 12640253, \\ 
12740133, 13640260).
VL is supported by JSPS Research for the Future Program
(No. JSPS-RFTF 97P01102).
SE, KN and JN are JSPS Research Fellows.

\end{document}